\documentclass[prb,aps,preprint]{revtex4}
\usepackage{amsmath}
\usepackage{amssymb}
\usepackage{amsbsy}
\usepackage{graphicx}
\usepackage{textcomp}

\newcommand{\unit}[2]%
{\mbox{\ensuremath{#1}}\mbox{\,\ensuremath{\mathrm{#2}}}}

\newcommand{\VSi}{V$_3$Si}
\newcommand{\Bctwo}{$B_{\mathrm{c}2}$}

\begin{document}
\title{Rotating sample magnetometer for cryogenic temperatures
and high magnetic fields}

\author{M.~Eisterer, F.~Hengstberger, C.~S.~Voutsinas, N.~H\"orhager,
S.~Sorta, J.~Hecher, and  H.~W.~Weber}
\email{eisterer@ati.ac.at}
\affiliation{Atominstitut,
Vienna University of Technology,
Stadionallee 2,
1020 Vienna,
Austria}

\begin{abstract}
We report on the design and implementation
of a rotating sample magnetometer (RSM)
operating in the variable temperature insert
of a cryostat equipped with a high-field magnet.
The limited space and the cryogenic temperatures
impose the most critical design parameters:
the small bore size of the magnet
requires a very compact pick-up coil system
and the low temperatures demand a very careful design of the bearings.
Despite these difficulties the RSM
achieves excellent resolution at high magnetic field sweep rates,
exceeding that of a typical vibrating sample magnetometer
by about a factor of ten.
In addition the gas-flow cryostat and the high-field superconducting magnet
provide a temperature and magnetic field range
unprecedented for this type of magnetometer.
\end{abstract}

\maketitle

\section{Introduction}
Among the instruments for measuring the magnetic moment of a sample
at high applied fields and cryogenic temperatures
two types of magnetometers are most common:
SQUID-based systems,
which detect the flux through a closed superconducting loop,
and the vibrating sample magnetometer\cite{Fon56} (VSM),
which measures the change of flux through normal conducting pick-up coils.
Although the resolution of SQUID magnetometers exceeds that
of a VSM by several orders of magnitude,
the measurement speed of the VSM has made this instrument
very popular for a fast characterisation of superconductors
or magnetic materials,
in particular,
if the resolution is not critical.

In a VSM the harmonic translation of the sample induces
a voltage in nearby pick-up coils,
which is proportional to the magnetic moment of the sample.
The obvious idea of rotating the sample,
has,
however,
received less attention
and only a limited number of instruments has been built.%
\cite{Fla70,Hud73,Moi84,Zaw06}
There is,
to our knowledge, 
no RSM capable of measuring at cryogenic temperatures
and high magnetic fields.
This is probably due to the challenging mechanical construction:
the bearing system must be compatible with cryogenic temperatures
and the small bore size of high field magnets
requires a very compact design,
which limits the radius of the sample rotation.

A simple consideration shows that despite this restriction
the sensitivity of an RSM should exceed that of a VSM.
The voltage induced in the pick-up coils
is the product of the sample's speed $v$
and a geometric factor $\Phi$,
which depends on the arrangement of the pick-up coils relative to the sample:
$U \propto v\,\Phi$.
At a given frequency $f$,
the speed of a sample rotating at a distance $R$ from the centre
is $v_{\mathrm{RSM}}=2\pi f R$
and the maximum speed of a sample oscillating with amplitude $A$ is
$v_{\mathrm{VSM}}=2\pi f A$,
i.e.,
the RSM sample speed is greater if $R>A$.
This condition is easily met also in the narrow sample space
of a cryostat equipped with a high-field magnet.
Moreover,
there is a great potential to enhance the geometric factor $\Phi$
by rotating the sample very close to the pick-up coils.

\section{Description of the RSM}

\begin{figure}
  \centering
  \includegraphics[width=0.35\textwidth]{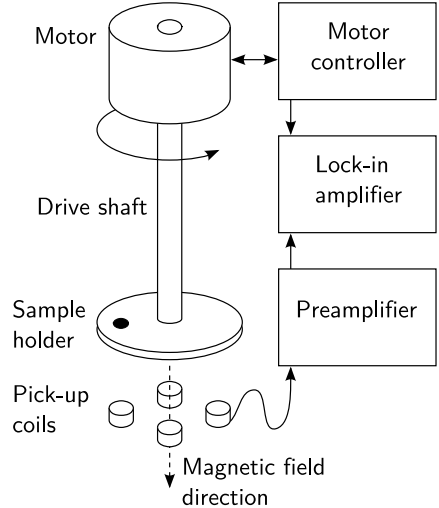}
  \caption{\label{fig:sketch}
    Schematic diagram of the RSM setup. 
    A motor outside the VTI
    connects to the drive shaft,
    which rotates the sample holder.
    The signal of the four pick-up coils
    is preamplified and measured by a lock-in amplifier
    using the reference signal provided by the motor controller.}
\end{figure}

Figure~\ref{fig:sketch} sketches the main components of the RSM setup,
which is similar to previously developed instruments
and installed vertically into a gas-flow cryostat.
The mode of operation is simple:
a motor rotates the sample,
which passes the pick-up coils positioned underneath.
The induced voltage is preamplified
and the harmonics are measured by a lock-in amplifier
using the reference signal provided by the motor controller.

We discuss the design of some crucial components in detail below.

\subsection{Mechanical Parts}

\begin{figure}
  \centering
  \includegraphics[width=.35\textwidth]{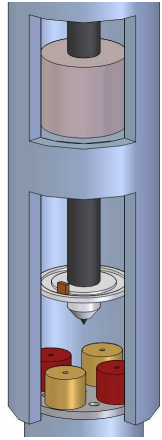}
  \caption{\label{fig:bottom}
    3D visualisation of the bottom part of the RSM.
    All components are inside a stainless steel tube
    coaxial with the magnetic field of the solenoid.
    The cylindrical bearings,
    which support the drive shaft
    are connected to the RSM tube
    (only the lowermost is shown).
    The sample (brown cube) is positioned in the ring-shaped grove
    of the aluminium sample holder
    and rotates closely above the coils.
    Here,
    the sample holder is shifted to show the bottom titanium tip (black cone)
    and the pick-up coil system
    (equal colours refer to equal polarity).}
\end{figure}

A stainless steel tube
of about one meter length and 28.5 mm in diameter
(2.5\,mm less than the inner diameter of our VTI)
contains the pick-up coil system
and all the mechanical components,
except the brushless DC motor,
which is outside the VTI at ambient atmosphere
and flanged to the top of the tube.
The motor axle connects through a vacuum-tight seal
to a short rod made of rubber.
This elastic intermediate piece,
which has a rigid connection with the drive shaft,
but is slideably mounted to the motor axle,
fulfils two functions:
it damps vibrations of the motor 
and compensates the differences in the thermal expansion
of the stainless steel tube and the aluminium drive shaft
(a few millimetres when cooling the RSM to liquid helium temperature).

The drive shaft
(6.3\,mm in diameter)
is coated with a carbon fibre reinforced polymer
and supported by three plain bearings made of graphalloy\cite{graphalloy} 
(the hollow cylinders have a length of 25\,mm and an outer diameter of 19\,mm). 
The aluminium sample holder (see Fig.~\ref{fig:bottom})
is screwed to the end of the drive shaft
and can be removed for sample mounting.
A ring-shaped groove machined close to the edge of the disk contains the sample
(usually glued with vacuum grease)
and reduces the sample to pick-up coil distance to about one millimetre.
The bottom bearing is the most crucial component,
because it has to guarantee stable rotation
(the main source of noise are vibrations)
and operate with low friction to avoid heating.
This is achieved by the titanium tip of the sample holder,
which spins in the conical hole of the metal disk underneath.
The instrument rotates at 15\,Hz,
because the signal-to-noise ratio is optimal at this speed 
and the heat generation acceptably low.
This corresponds to a signal with a frequency of 30\,Hz,
because the period of the pick-up coil system is half a rotation
(see Fig.~\ref{fig:signal}).

The cooling power is provided by the helium gas
flowing through holes at the bottom of the tube.
(We tested the instrument also successfully in liquid helium.)
Because the temperature sensor cannot be mounted to the rotating parts,
it is positioned at a small distance just above the sample.

\subsection{Pick-up Coil System}

\begin{figure}
  \centering
  \includegraphics[width=.5\textwidth]{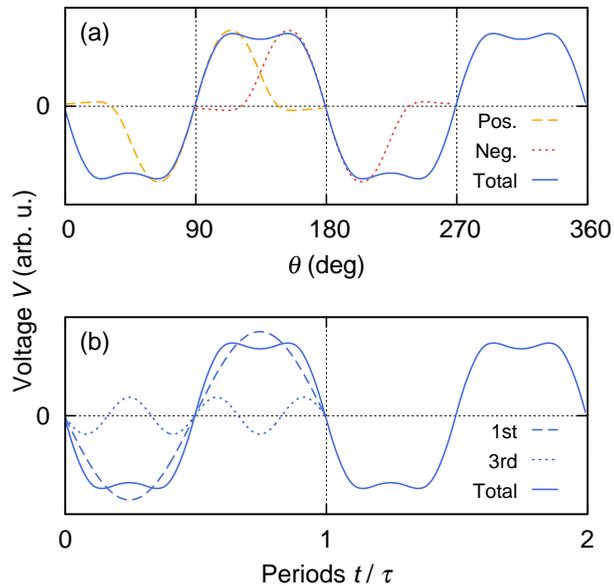}
  \caption{\label{fig:signal}
    Signal of the pick-up coil system.
    (a)~Four coils with alternating polarity
    (see Fig.~\ref{fig:bottom})
    compose the total signal.
    The graph also shows the contribution of two neighbouring coils
    positioned at 90 and 180$^\circ$
    to the total signal.
    This arrangement has two effects:
    first,
    an external field sweep generates no net signal,
    second,
    the individual signals interfere positively
    and create strong peaks.
    (b)~The period of the system is 180$^\circ$,
    i.e.,
    half a rotation.
    The first harmonic is the principle component
    and its peaks are shaped by the third harmonic
    (see also Fig.~\ref{fig:radius}).
    All higher harmonics are almost negligible.}
\end{figure}

The pick-up coil system is a fundamental part of the instrument,
because the geometry factor $\Phi$
defines the signal (together with the speed $v$).
It is therefore necessary to analyse
the coil parameters in detail.

Earlier work\cite{Zaw06}
aimed at increasing the signal by maximising the flux through a single coil
and suggested to use coils with a cross-section (perpendicular to the axis),
which is similar to that of the sample.
We do not follow this design criterion,
because a careful analysis requires that
instead of the magnetic flux the measured quantity,
i.e.,
the strongest harmonic of the induction voltage,
has to be optimised.
In order to calculate the induced voltage,
the full time dependence of the flux must be known;
calculating the flux only at a certain position
(above the centre of the coil)
is not sufficient.

Our numerical calculations
simulate the movement of a permanent magnet
($3\times3\times3$\,mm$^3$)
over the pick-up coil system.
The radius of the innermost winding $r_\mathrm{i}=1\,\mathrm{mm}$
and the sample-to-coil distance $d=1\,\mathrm{mm}$ are fixed,
whereas the outer radius $r_\mathrm{o}$
and the height $h$ of the coils
are free parameters in the simulations.
We limited the calculations to a set of four coils,
symmetrically positioned at a distance of 8\,mm from the centre.
A higher number of coils is impractical,
because mounting and wiring four coils is already
a very tedious task in the narrow sample space of the RSM.
The polarity of the coils alternates as displayed in Fig.~\ref{fig:bottom},
which eliminates the induction produced by the external field sweep.%
\footnote{The signal of the two coil pairs
(opposite coils are connected in series)
deviates by about one percent
when sweeping the external magnetic field.
This deviation stems from imperfections in the windings 
and is nulled at the preamplification stage.}
After computing the flux as a function of the rotation angle 
by integrating the magnetic field through each winding 
and summing over all turns
we calculate the induced voltage by derivation
(the time and the angular derivative are directly proportional) 
and extract the harmonics by Fourier transformation
(see Fig.~\ref{fig:signal} for an example).
Our simulation exactly represents the pick-up coil system
without any simplifications,
except that larger wire diameters of up to 100\,\textmu{}m
have been used to speed up the calculations.

\begin{figure}
  \centering
  \includegraphics[width=.5\textwidth]{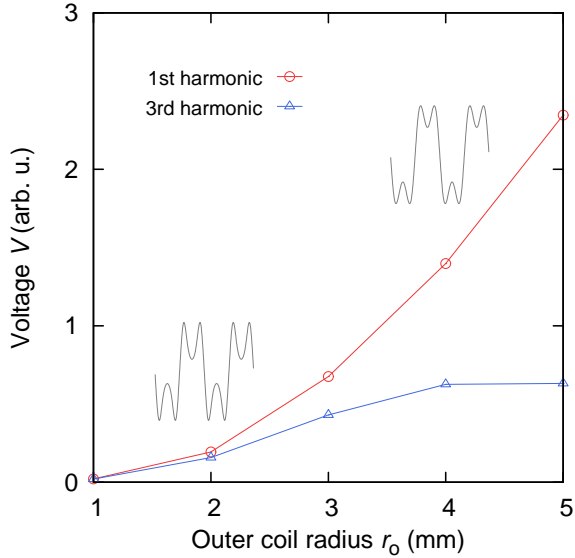}
  \caption{\label{fig:radius}
    Dependence of the pick-up coil signal on the outer coil radius.
    The magnitude of the first harmonic,
    which is the measurement quantity,
    does not decrease if the 
    coil cross-section $r_0^2\pi$ exceeds that of the sample
    (approximately at $r=2\,\mathrm{mm}$).
    The derivative of the first harmonic with respect
    to $r_0$ still increases at $r_0=4\,\mathrm{mm}$,
    which demonstrates that the outermost windings
    contribute most to the signal.
}
\end{figure}

The calculations,
which were carried out by varying $r_\mathrm{o}$ for various $h$,
show that the total signal is mainly composed
of the first and the third harmonic
(see Fig.~\ref{fig:signal})
and that the reduction of the third harmonic results from the overlap
of the coil signals.
We find that the measured quantity,
i.e.,
the strong first harmonic of the induction voltage,
increases up to the maximum outer coil radius
compatible with our setup ($r_0=4\,\mathrm{mm}$),
even if the coil cross-section exceeds that of the sample.
(Figure~\ref{fig:radius} presents this calculation
with the coil height of the final setup.)
This highlights the importance of exact numerical calculations
of the true signal,
because the application of a simpler design criterion
based on the magnetic flux\cite{Zaw06}
would result in coils with smaller radii and lower signal.
Moreover,
the fact that the derivative of the first harmonic with respect
to $r_0$ still increases at $r_0=4\,\mathrm{mm}$
shows that the outermost windings contribute most to the signal.

Because space and not a loss in signal strength
is the limiting factor in our configuration,
we use the largest possible pick-up coils
compatible with our setup:
about 6\,mm high with a radius of 4\,mm
having 8000 turns of 38\,\textmu{}m thick copper wire.

\section{Performance}

\subsection{Calibration}
In a VSM the distance of the sample to the pick-up coils
is usually much larger than its dimensions,
which allows to treat the sample as a dipole.
The small sample to pick-up coil distance of the RSM
enhances the resolution,
but makes the instrument also sensitive to the geometry of the sample.
(Note,
that this is not a demagnetisation effect,
which depends on the stray field of the sample
and is independent of the measurement method.)
It is therefore necessary to calibrate the RSM 
for every sample geometry and a reference measurement
in another magnetometer is needed
to determine the calibration constant.

\subsection{Background, Noise and Resolution}

\begin{figure}
  \centering
  \includegraphics[width=.5\textwidth]{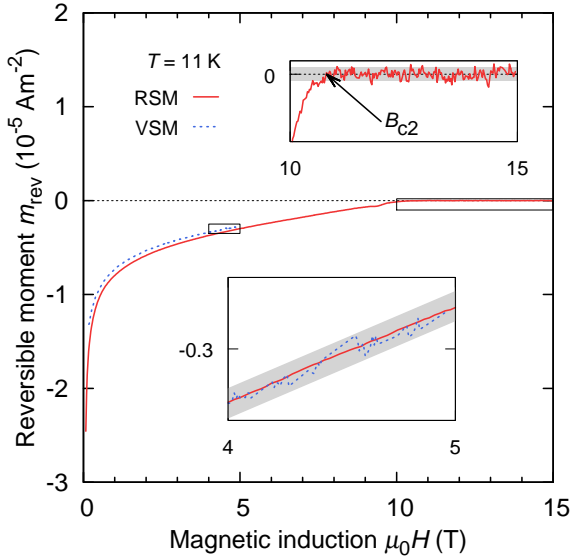}
  \caption{\label{fig:comp}
    Comparison of VSM and RSM measurements
    on a \VSi{} single crystal after background subtraction.
    The RSM measurement shows very good agreement 
    with the VSM data up to the maximum field of the VSM
    (the RSM data is shifted for clarity in this graph only).
    The lower inset demonstrates the superior resolution of the RSM
    (the grey area corresponds to the typical VSM noise of
    $\pm 10^{-7}$\,Am$^{-2}$).
    The upper inset shows the determination
    of the upper critical field \Bctwo{} of the superconductor
    (at this field the magnetic moment of the sample vanishes).
    Here,
    the grey area indicates the RSM noise,
    which is as low as $10^{-8}$\,Am$^{-2}$ at 15\,T.
    This allows a precise determination
    of this parameter up to high magnetic fields.}
\end{figure}

Measurements without a sample show
that the background of the sample holder
increases with the magnetic field
and is as low as
$5\times10^{-8}\,\mathrm{A}/\mathrm{m}^2$ at 15\,T.
The low background is a consequence of the sample rotation:
contrary to a VSM,
where the oscillation moves the sample
out of the centre of the magnet to lower fields
the sample rotates around the symmetry axis of the magnet in the RSM
and the magnetic field inhomogeneity does not have an effect
(also the sample holder does not break the symmetry).
All the measurements presented herein use the first harmonic,
because experiments have shown that
it has the highest signal-to-noise ratio
(about a factor of ten less for the third harmonic).

The noise decreases by orders of magnitude
when disconnecting the motor and the drive shaft
by removing the intermediate piece.
This demonstrates that the main source of noise
are mechanical vibration induced by the rotating drive shaft.
Under the assumption that the vibrations
move the pick-up coil system
in radial direction out of the centre of the magnet
we can estimate the amplitude of these vibrations
from the pick-up coil inhomogeneity
and the radial magnetic field gradient.
We find a vibration amplitude of about 5\,$\mu$m.

To determine the resolution
we carried out measurements
on a superconducting \VSi{} single crystal
($0.5 \times 0.5 \times 3$\,mm$^3$)
in the RSM and in our Oxford Maglab VSM
($1.5$\,mm amplitude, 55\,Hz)
using a field sweep rate of 0.5\,$\mathrm{T}/\mathrm{min}$.
Here,
the background,
i.e.,
the non-superconducting contribution,
is due to the paramagnetic signal of the \VSi{} sample and is highly linear.
Figure~\ref{fig:comp}
shows the reversible magnetisation after subtracting this term 
and demonstrates that both instruments agree well
up to the maximum field of the VSM.
The low noise level of the RSM,
which is more than a factor of ten lower than that of the VSM,
is compatible with the resolution of
a magnetic moment as low as $10^{-8}$\,Am$^{-2}$
at a background field of 15\,T.
This allows to determine,
for example,
the upper critical field \Bctwo{}
(the field,
where the magnetic moment of the superconductor goes to zero)
with great precision.

The resolution of the RSM is excellent at high field sweep rates
and can be further enhanced at the expense of measurement time
by reducing the field sweep rate
and increasing the averaging time of the lock-in amplifier.

\section{Summary}
We successfully implemented a rotating sample magnetometer
operating at cryogenic temperatures and high magnetic fields.
After calibration the RSM measurements compare well to VSM data.
The resolution of the RSM is $10^{-8}$\,Am$^{-2}$ at 15\,T
(exceeding that of a typical VSM by about a factor of ten)
and the background signal of the sample holder is
is as low as $5\times10^{-8}$\,Am$^{-2}$ at 15\,T.
The magnetometer is thus ideally suited for a precise and swift characterisation
of superconducting and magnetic samples in a wide temperature
and magnetic field range.

\section*{Acknowledgements}
We wish to express our sincere gratitude to H. Hartmann for technical support
and acknowledge the assistance of J. Emhofer with the CAD software.


\end{document}